# CURRENT STUDIES ON INTRUSION DETECTION SYSTEM, GENETIC ALGORITHM AND FUZZY LOGIC

Mostaque Md. Morshedur Hassan

LCB College, Maligaon, Guwahati, Assam, India.

`mostaq786@gmail.com`

## *ABSTRACT*

*Nowadays Intrusion Detection System (IDS) which is increasingly a key element of system security is used to identify the malicious activities in a computer system or network. There are different approaches being employed in intrusion detection systems, but unluckily each of the technique so far is not entirely ideal. The prediction process may produce false alarms in many anomaly based intrusion detection systems. With the concept of fuzzy logic, the false alarm rate in establishing intrusive activities can be reduced. A set of efficient fuzzy rules can be used to define the normal and abnormal behaviors in a computer network. Therefore some strategy is needed for best promising security to monitor the anomalous behavior in computer network. In this paper I present a few research papers regarding the foundations of intrusion detection systems, the methodologies and good fuzzy classifiers using genetic algorithm which are the focus of current development efforts and the solution of the problem of Intrusion Detection System to offer a real-world view of intrusion detection. Ultimately, a discussion of the upcoming technologies and various methodologies which promise to improve the capability of computer systems to detect intrusions is offered.*

## 1. INTRODUCTION

Intrusion detection is the process of monitoring the events ([1], [2], [3]) occurring in a computer system or network and analyzing them for signs of probable incidents, which are violations or forthcoming threats of violation of computer security strategies, adequate used policies, or usual security practices. Intrusive events to computer networks are expanding because of the liking of adopting the internet and local area networks [4] and new automated hacking tools and strategy. Computer systems are evolving to be more and more exposed to attack, due to its wide spread network connectivity.

Currently, networked computer systems play an ever more major role in our society and its economy. They have become the targets of a wide array of malicious threats that invariably turn into real intrusions. This is the reason computer security has become a vital concern for network practitioner. Too often, intrusions cause disaster inside LANs and the time and cost to renovate the damage can grow to extreme proportions. Instead of using passive measures to repair and patch security hole once they have been exploited, it is more efficient to take up a proactive measure to intrusions.

Intrusion Detection Systems (IDS) are primarily focused on identifying probable incidents, monitoring information about them, tries to stop them, and reporting them to security administrators [5] in real-time environment, and those that exercise audit data with some delay (non-real-time). The latter approach would in turn delay the instance of detection. In addition, organizations apply IDSs for other reasons, such as classifying problems with security policies, documenting existing attacks, and preventing individuals from violating security policies. IDSs have become a basic addition to the security infrastructure of almost every organization. A usual Intrusion Detection System is demonstrated in Figure 1.

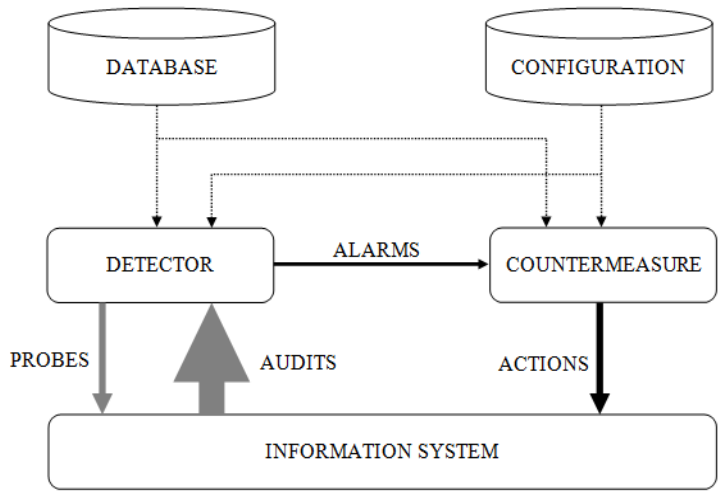

NOTE: The arrow lines symbolize the amount of information flowing from one component to another

Figure 1. Very Simple Intrusion Detection System

Intrusion Detection Systems are broadly classified into two types. They are host-based and network-based intrusion detection systems. Host-based IDS employs audit logs and system calls as its data source, whereas network-based IDS employs network traffic as its data source. A host-based IDS consists of an agent on a host which identifies different intrusions by analyzing audit logs, system calls, file system changes (binaries, password files, etc.), and other related host activities. In network-based IDS, sensors are placed at strategic position within the network system to capture all incoming traffic flows and analyze the contents of the individual packets for intrusive activities such as denial of service attacks, buffer overflow attacks, etc. Each approach has its own strengths and weaknesses. Some of the attacks can only be detected by host-based or only by network-based IDS.

The two main techniques used by Intrusion Detection Systems for detecting attacks are Misuse Detection and Anomaly Detection. In a misuse detection system, also known as signature based detection system; well known attacks are represented by signatures. A signature is a pattern of activity which corresponds to intrusion. The IDS identifies intrusions by looking for these patterns in the data being analyzed. The accuracy of such a system depends on its signature database. Misuse detection cannot detect novel attacks as well as slight variations of known attacks.

An anomaly-based intrusion detection system inspects ongoing traffic, malicious activities, communication, or behavior for irregularities on networks or systems that could specify an attack. The main principle here is that the attack behavior differs enough from normal user behavior that International Journal of Distributed and Parallel Systems (IJDPS) Vol.4, No.2, March 2013it cannot be detected by cataloging and identifying the differences involved. By creating supports of standard behavior, anomaly-based IDS can view when current behaviors move away statistically from the normal one. This capability gives the anomaly-based IDS ability to detect new attacks for which the signatures have not been created. The main disadvantage of this method is that there is no clear cut method for defining normal behavior. Therefore, such type of IDS can report intrusion, even when the activity is legitimate.

One of the major problems encountered by IDS is large number of false positive alerts that is the alerts that are mistakenly analyzed normal traffic as security violations. An ideal IDS does not produce false or inappropriate alarms. In practice, signature based IDS found to produce more false alarms than expected. This is due to the very general signatures and poor built in verification tool to authenticate the success of the attack. The large amount of false positives in the alert logs generates the course of taking corrective action for the true positives, i.e. delayed, successful attacks, and labor intensive.

My goal is to detect novel attacks by unauthorized users in network traffic. I consider an attack to be novel if the vulnerability is unknown to the target's owner or administrator, even if the attack is generally known and patches and detection tests are available. I mostly like to cite four types of remotely launched attacks: denial of service (DOS), U2R, R2L, and probes. A DoS attack is a type of attack in which the unauthorized users build a computing or memory resources too busy or too full to provide reasonable networking requests and hence denying users access to a machine e.g. ping of death, neptune, back, smurf, apache, UDP storm, mail bomb etc. are all DoS attacks. A remote to user (U2R) attack is an attack in which a user forwards networking packets to a machine through the internet, which he/she does not have right of access in order to expose the machines vulnerabilities and exploit privileges which a local user would have on the computer e.g. guest, xlock, xnsnoop, sendmail dictionary, phf etc. A R2L attacks are regarded as the exploitations in which the unauthorized users start off on the system with a normal user account and tries to misuse vulnerabilities in the system in order to achieve super user access rights e.g. xterm, perl. A probing is an attack in which the hacker scans a machine or a networking device in order to determine weaknesses or vulnerabilities that may later be exploited so as to negotiate the system. This practice is commonly used in data mining e.g. portsweep, saint, mscan, nmap etc.

The Intrusion Detection System (IDS) is also carried out by implementing Genetic Algorithm (GA) to efficiently identify various types of network intrusions. The genetic algorithm [1] is applied to achieve a set of classification rules from the support-confidence framework, and network audit data is employed as fitness function to judge the quality of each rule. The created rules are then used to classify or detect network intrusions in a real-time framework. Unlike most available GA-based approaches remained in the system, because of the easy demonstration of rules and the efficient fitness function, the proposed system is very simple to employ while presenting the flexibility to either generally detect network intrusions or precisely classify the types of attacks.

The normal and the abnormal intrusive activities in networked computers are tough to forecast as the boundaries cannot be well explained. This prediction process may generate false alarms [1] in many anomaly based intrusion detection systems. However, with the introduction of fuzzy logic, the false alarm rate in determining intrusive activities can be minimized; a set of fuzzy rules (non-crisp fuzzy classifiers) can be employed to identify the normal and abnormal behavior in computer networks, and fuzzy inference logic can be applied over such rules to determine when an intrusion is in progress. The main problem with this process is to make good fuzzy classifiers to detect intrusions.

It has been shown by Baruah [6] that a fuzzy number [a, b, c] is concluded with reference to a membership function μ(x) remaining within the range between 0 and 1, a ≤ x ≤ c. Further, he has extended this definition in the following way. Let µ1(x) and µ2(x) be two functions, $0 \leq \mu_2(x) \leq \mu_1(x) \leq 1$. He has concluded µ1(x) the fuzzy membership function, and µ2(x) a reference function, such that (µ1(x) – µ2(x)) is the fuzzy membership value for any x. Finally he has characterized such a fuzzy number by {x, µ1(x), µ2(x); x ∈ Ω}.

The complement of $\mu_x$ is always counted from the ground level in Zadehian's theory [9], whereas it actually counted from the level if it is not as zero that is the surface value is not always zero. If other than zero, the problem arises and then we have to count the membership value from the surface for the complement of $\mu_x$.

In Figure 2, Baruah [6] explained that for a fuzzy number A = [a, b, c], the value of membership for any x ∈ Ω is specified by μ(x) for a ≤ x ≤ c, and is taken as zero otherwise. For the fuzzy number $A^C$, the value of membership for any x ∈ Ω is given by (1 - μ(x)) for a ≤ x ≤ c, otherwise, the value is 1. The main difference is that for $A^C$ the membership function holds 1 all over the place with the reference function being μ(x), whereas for A the membership function is μ(x) with the reference function being 0 everywhere.

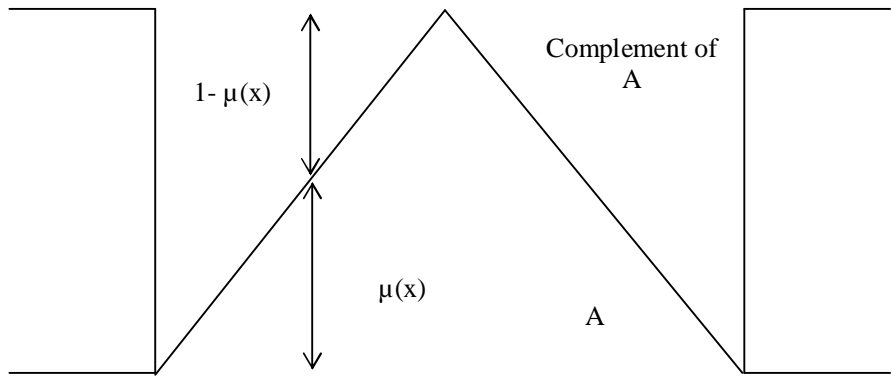

Figure 2. Extended definition of Fuzzy Set

This proposed system forwarded a definition of complement of an extended fuzzy set in which the fuzzy reference function is not always taken as zero. The definition of complement of a fuzzy set forwarded by Baruah ([6], [7]), Neog and Sut [8] could be viewed a particular case of what I am proposing. I would use Baruah's definition of the complement of a normal fuzzy set in my work.

## 2. INTRUSION DETECTION STRATEGIES

The intrusion detection strategies concerns four primary issues. First is the dataset that is captured from network communications. The second is Genetic Algorithms (GA) which use mutation, recombination, and selection applied to a population of individuals in order to evolve iteratively better and better solutions and a way to generate fuzzy rules to characterize normal and abnormal behavior of network systems. The third is to generate alerts and reports for malicious traffic behavior, and the fourth is the maintenance of the ids for observation of placement of sensors, and qualified trained intrusion analysts so that the latest malicious traffic is being detected.

## 2.1. The Dataset

To implement the algorithm and to evaluate the performance of the system, I propose the standard datasets employed in KDD Cup 1999 "Computer Network Intrusion Detection" competition.

The KDD 99 intrusion detection datasets depends on the 1998 DARPA proposal, which offers designers of intrusion detection systems (IDS) with a standard on which to evaluate different methodologies ([21], [24]). Hence, a simulation is being prepared from a fabricated military network with three 'target' machines running various services and operating systems. They also applied three extra machines to spoof different IP addresses for generating network traffic.

A connection is a series of TCP packets beginning and ending at some well defined periods, between which data floods from a source IP address to a target IP address under some well defined protocol ([21], [22], [24]). It results in 41 features for each connection.

Finally, there remains a sniffer that accounts all network traffic by means of the TCP dump format [24]. The total simulated period is seven weeks. Normal connections are shaped to outline that expected in a military network and attacks are categorized into one of four types: User to Root; Remote to Local; Denial of Service; and Probe.

The KDD 99 intrusion detection benchmark consists of different components [23]:

kddcup.data; kddcup.data_10_percent; kddcup.newtestdata_10_percent_unlabeled;

kddcup.testdata.unlabeled; kddcup.testdata.unlabeled_10_percent; corrected.

I propose to use "kddcup.data_10_percent" as training dataset and "corrected" as testing dataset. In this case the training set consists of 494,021 records among which 97,280 are normal connection records, while the test set contains 311,029 records among which 60,593 are normal connection records. Table 1 shows the intrusion types distribution in the training and the testing datasets.

Table 1. Intrusion types distribution in datasets

| Dataset | normal | prob | Dos | u2r | r2l | Total |
|---|---|---|---|---|---|---|
| Train | 97280 | 4107 | 391458 | 52 | 1124 | 494021 |
| Test ("corrected") | 60593 | 4166 | 229853 | 228 | 16189 | 311029 |

## 2.2. Genetic algorithm

### 2.2.1. Genetic algorithm overview

A Genetic Algorithm (GA) is a programming technique that uses biological evolution as a problem solving strategy [20]. It is based on Darwinian's theory of evolution and survival of fittest to make effective a population of candidate result near a predefined fitness [13].

The proposed GA based intrusion detection system holds two modules where each acts in a dissimilar stage. In the training stage, a set of classification rules are produced from network audit data using the GA in an offline background. In the intrusion detection phase, the generated rules are employed to classify incoming network connections in the real-time environment. Once the rules are generated, the intrusion detection system becomes simple, experienced and efficient one.

GA applies an evolution and natural selection that employs a chromosome-like data structure and evolve the chromosomes by means of selection, recombination and mutation operators [13]. The process generally starts with randomly generated population of chromosomes, which signify all possible solution of a problem that are measured candidate solutions. From each chromosome different positions are set as bits, characters or numbers. These positions are regarded as genes. An evaluation function is employed to find the decency of each chromosome according to the required solution; this function is known as "Fitness Function". During the process of evaluation "Crossover" is applied to have natural reproduction and "Mutation" is applied to mutation of species [13]. For survival and combination the selection of chromosomes is partial towards the fittest chromosomes.

When I use GA for solving various problems three factors will have crucial impact on the use of the algorithm and also of the applications [2]. The factors are : i) the fitness function, ii) the representation of individuals, and iii) the genetic algorithm parameters. The determination of these factors often depends on implementation of the system.

### 2.2.2 Fuzzy logic

Zadeh explained that Fuzzy logic [9] is an extension of Boolean logic that is often used for computer-based complex decision making. While in classical Boolean logic an element can be either a full member or non-member of a Boolean (sometimes called "crisp") set, the membership of an element to a fuzzy set can be any value within the interval [0, 1], allowing also partial membership of an element in a set.

A fuzzy expert system consists of three different types of entities: fuzzy sets, fuzzy variables and fuzzy rules. The membership of a fuzzy variable in a fuzzy set is determined by a function that produces values within the interval [0, 1]. These functions are called membership functions. Fuzzy variables are divided into two groups: antecedent variables, that are assigned with the input data of the fuzzy expert system and consequent variables, that are assigned with the results computed by the system.

The fuzzy rules determine the link between the antecedent and the consequent fuzzy variables, and are often defined using natural language linguistic terms. For instance, a fuzzy rule can be "if the temperature is cold and the wind is strong then wear warm clothes", where temperature and wind are antecedent fuzzy variables, wear is a consequent fuzzy variable and cold, strong and warm clothes are fuzzy sets.

The process of a fuzzy system has three steps. These steps are Fuzzification, Rule Evaluation, and Defuzzification. In the fuzzification step, the input crisp values are transformed into degrees of membership in the fuzzy sets. The degree of membership of each crisp value in each fuzzy set is determined by plugging the value into the membership function associated with the fuzzy set. In the rule evaluation step, each fuzzy rule is assigned with a strength value. The strength is determined by the degrees of memberships of the crisp input values in the fuzzy sets of antecedent part of the fuzzy rule. The defuzzification stage transposes the fuzzy outputs into crisp values.

It has been revealed by Baruah [6] that a fuzzy number [a, b, c] can be explained with reference to a membership function $\mu(x)$ remaining between 0 and 1, $a \leq x \leq c$. Further, he has extended this definition in the following way. Let $\mu_1(x)$ and $\mu_2(x)$ be two functions, $0 \leq \mu_2(x) \leq \mu_1(x) \leq 1$. He has concluded $\mu_1(x)$ the fuzzy membership function, and $\mu_2(x)$ a reference function, such that $(\mu_1(x) - \mu_2(x))$ is the fuzzy membership value for any x. Finally he has characterized such a fuzzy number by $\{x, \mu_1(x), \mu_2(x); x \in \Omega\}$.

The complement of $\mu_x$ is always counted from the ground level in Zadehian's theory [9], whereas it actually counted from the level if it is not as zero that is the surface value is not always zero. If other than zero, the problem arises and then we have to count the membership value from the surface for the complement of $\mu_x$. Thus I could conclude the following statement –

Complement of $\mu_x$ = 1 for the entire level

Membership value for the complement of $\mu_x$ = 1- $\mu_x$

I have forwarded Baruah's definition of complement of an extended fuzzy set where the fuzzy reference function is not always taken as zero. The definition of complement of a fuzzy set recommend by Baruah ([6], [7]), Neog and Sut [8] could be considered a particular case of what I am giving. I would use Baruah's definition of the complement of a normal fuzzy set in my proposed work.

In the two classes' classification problem, two classes are available where every object should be classified. These classes are called positive (abnormal) and negative (normal). The data set employed by the learning algorithms holds a set of objects where each object contains n+1 attributes. The first n attributes identifies the monitored parameters of the object characteristics and the last attribute identifies the class where the object belongs to the classification attribute.
A fuzzy classifier is a set of two rules for solving the two classes' classification problem, one for the normal class and other for the abnormal class, where the conditional part is described by means of only the monitored parameters and the conclusion part is viewed as an atomic expression for the classification attribute.

### 2.2.3 Fitness function

The authors in [1] used the fuzzy confusion matrix to calculate the fitness of a chromosome. In the fuzzy confusion matrix the fuzzy truth degree of the condition represented by the chromosome and the fuzzy negation operator are used directly. The fitness of a chromosome for the abnormal class is evaluated according to the following set of equations:

$$TP = \sum_{i=1}^{p} \text{predicted (class\_data}_i\text{)}$$

$$TN = \sum_{i=1}^{q} 1 - \text{predicted (other\_class\_data}_i\text{)}$$

$$FP = \sum_{i=1}^{q} \text{predicted (other\_class\_data}_i\text{)}$$

$$FN = \sum_{i=1}^{p} 1 - \text{predicted (class\_data}_i\text{)}$$

Where,

Sensitivity = TP/(TP+FN)
Specificity = FP/( FP+ TN)

Length = 1 – chromosome_length/10

So finally Fitness of a chromosome is calculated as follows –

Fitness = W1 * Sensitivity + W2 * Specificity + W3 * Length

Where,

TP, TN, FP, FN are true positive, true negative, false positive, false negative value for the rule, p is the number of samples of the evolved class in the training data set, q is the number of samples of the remaining class in the training data set, predicted is the fuzzy value of the conditional part of the rule, class_data$_i$ is an element of the subset of the training samples of the evolved class, other_class_data$_i$ is an element of the subset of the remaining classes in the training samples, and W1, W2, W3 are the assigned weights for each rule characteristics respectively.

## 2.3 Generate alerts and reports

The reports portraits an entire image of the status of the network under observation. It handles all the output from the system, whether that be an automated response to the suspicious activity, or which is most common, the notification of some security officer. IDSs should provide facilities for practitioners to fine-tune thresholds for generating alarms as well as facilities for suppressing alarms selectively.

Reporting can demonstrate the economic value of the monitoring tools. It can also ease the burden of monitoring. The IDS should generate reports that help practitioners investigate the alarms. In addition, the intrusion detection system can assist network practitioners prioritize their tasks, by assigning priorities to alarms, or providing each alarm to a practitioner for further investigation.

## 2.4 IDS Maintenance

### 2.4.1 Maintenance

IDS maintenance is essential for all IDS technologies because all sorts of threats and prevention technologies are constantly varying, patches, signatures, and configurations must be modernized to ensure that the latest malicious traffic is being detected and prevented. We could maintain IDS from a console using a graphical user interface (GUI), application, or secure web-based interface. Network administrators could monitor IDS components from the console to make sure they are operational, validate they are working properly, and carry out vulnerability assessments (VA) and updates.

### 2.4.2 Tuning

To be effective in detection policy, IDS must be tuned precisely. Tuning requires varying different settings to be in conformity with the security guiding principles and objective of the IDS administrator. Scanning techniques, thresholds, and focus can be regulated to make certain that anIDS is making out relevant data without overloading the network administrator with warnings or too many false positives. Tuning is time-consuming, but it must be performed to make sure an efficient IDS configuration. It is to be noted that tuning must be specific to the IDS product only.

**2.4.3 Detection Accuracy**

The accuracy of intrusion detection system relies on the technique in which it identifies, such as by the rule set. Signature-based detection detects only simple and recognized attacks, while anomaly-based detection can detect more types of attacks, but has a higher number of false positives ratios. Tuning is essential to reduce the number of false positives and to make the data further functional.

## 3. CHALLENGES IN IDS

There are number of challenges that impact on organization's decision to use IDS. In this section I have described a few challenges that the organizations encounter while installing an intrusion detection system. These are discussed below –

1. **Human intervention** - IDS technology itself is experiencing a lot of enhancements. It is therefore very important for organizations to clearly define their prospect from the IDS implementation. Till now IDS technology has not achieved a level where it does not require human interference. Of course today's IDS technology recommends some automation like reporting the administrator in case of detection of a malicious activity, avoiding the malicious connection for a configurable period of time, dynamically changing a router's access control list in order to prevent a malicious connection etc.
   Therefore the security administrator must investigate the attack once it is detected and reported, determine how it occurred, correct the problem and take necessary action to prevent the occurrence of the same attack in future.

2. **Historical analysis** - It is still very important factor to monitor the IDS logs regularly to continue on top of the incidence of events. Monitoring the logs on a daily basis is necessary to analyze the different type of malicious activities detected by the IDS over a period of time. Today's IDS has not yet achieved the level where it can provide historical analysis of the intrusive activities detected over a span of time. This is still a manual activity.

   Hence it is vital for an organization to have a distinct incident handling and response plan if an intrusion is detected and reported by the IDS. Also, the organization should have expert security personnel to handle this kind of situation.

3. **Deployment** - The success of an IDS implementation depends to a large degree on how it has been deployed. A lot of plan is necessary in the design as well as the implementation phase. In most cases, it is required to apply a fusion solution of network based and host based IDS to gain from both cases. In fact one technology complements the other.

   However, this decision can differ from one organization to another. A network based IDS is an instant choice for many organizations because of its capability to monitor multiple systems and also the truth that it does not need a software to be loaded on a production system different from host based IDS.

   Some organizations implement a hybrid solution. Organizations installing host based IDS solution needs remember that the host based IDS software is processor and memory challenging. So it is very important to have sufficient available resources on a system before establishing a host based sensor on it.

4. **Sensors** - It is important to maintain sensor to manager ratio. There is no strict rule as such for calculating this ratio. To a large degree it depends upon how many different types of traffic is monitored by each sensor and in which background. Most of the organizations deploy a ratio of 10:1, while some organizations maintain 20:1 and some others go for 15:1.

   It is very important to plan the baseline strategy before starting the IDS implementation and avoid false positives. A poorly configured IDS sensor may post a lot of false positive ratios to the console and even a ratio of 10:1 or even enough better sensors to the console ratio can be missing.

5. **False positive and negative alarms rate –** It is impossible for IDS to be ideal mostly because network traffic is so complicated. The erroneous results in IDS are divided into two types: false positives and false negatives. False positives take place when the IDS erroneously identify a problem with benign traffic. False negatives occur when redundant traffic is overlooked by the IDS. Both create problems for security administrators or practitioners and demands that the malicious threats must be detected powerfully. A greater number of false positives are generally more acceptable but can burden a security administrator with bulky amounts of data to filter through. However, because it is unnoticed, false negatives do not provide a security administrator a chance to check the data. Therefore IDS to be implemented should minimize both false positive and negative alarms.

6. **Signature database -** A common policy for IDS in detecting intrusions is to remember signatures of known attacks. The inherent weak points in relying on signatures are that the signature patterns must be acknowledged first. New threats are often unrecognizable by eminent and popular IDS. Signatures can be masked as well. The ongoing event between new attacks and detection systems has been a challenge. Therefore the signature database must be updated whenever a different kind of attack is detected and repair for the same is available.

7. **Monitor traffic in large networks** - Network Intrusion Detection System (NIDS) components are spotted throughout a network, but if not placed tactically, many attacks can altogether avoid NIDS sensors by passing through alternate ways in a network. Moreover, though many IDS products available in the market are efficient to distinguish different types of attacks, they may fail to recognize attacks that use many attack sources. Many IDS cannot cleverly correlate data from numerous sources. Newer IDS technologies must influence integrated systems to increase an overview of distributed intrusive activity. Therefore IDS must be able to successfully monitor traffic in a large network.

## 4. PRIOR WORKS

In this section, I describe the important and relevant research works of different authors that I have come across during the literature survey of my proposed work. I illustrate each attack manner and point to the impact of this attack and its intrusive activities. From an intruder's point of view, I analyze each of the attack's modes, intention, benefits and suitable conditions and try to find out the solution how to improve the attack by introducing the concept of fuzzy logic-based technique and genetic algorithm.

The normal and abnormal behaviors [1] in networked computers are hard to forecast, as the limits cannot be explained clearly. This prediction method usually generates fake alarms in many anomaly based intrusion detection systems.

In [1] the authors introduced the concept of fuzzy logic to reduce the fake alarm rate in determining intrusive behavior. The set of fuzzy rules is applied to identify the normal and abnormal behavior in a computer network. The authors proposed a technique to generate fuzzy rules that are able to detect malicious activities and some specific intrusions. This system presented a novel approach for the presentation of generated fuzzy rules in classifying different types of intrusions.

The advantage of their proposed mechanism is that the fuzzy rules are able to detect the malicious activities.

But they failed to implement the real time network traffic, more attributes for the classification rules. In determining the fuzzy rules, they used the concept of fuzzy membership function and reference function, but they said that the membership function and reference function are same. In reality, these two concepts are totally different concepts. I have forwarded the extended definition of fuzzy set of Baruah ([6], [7]), Neog and Sut [8].

In [9] Zadeh initiated the idea of fuzzy set theory and it was mainly intended mathematically to signify uncertainty and vagueness with formalized logical tools for dealing with the vagueness connected in many real world problems. The membership value to a fuzzy set of an element describes a function called membership function where the universe of discourse is the domain and the interval lies in the range [0, 1]. The value 0 means that the element is not a member of the fuzzy set; the value 1 means that the element is fully a member of the fuzzy set. The values that remain between 0 and 1 distinguish fuzzy members, which confined to the fuzzy set merely partially.

But the author gave an explanation that the fuzzy membership value and fuzzy membership function for the complement of a fuzzy set are same concepts and the surface value is always counted from the ground level.

Baruah ([6], [7]), Neog and Sut [8] have forwarded an extended definition of fuzzy set which enables us to define the complement of a fuzzy set. My proposed system agrees with them as this new definition satisfies all the properties regarding the complement of a fuzzy set.

In [2] Gong, Zulkernine, Abolmaesumi gave an implementation of genetic based approach to Network Intrusion Detection using genetic algorithm and showed software implementation to detect the malicious activities. The approach derived a set of classification rules from network audit data and utilizes a support-confidence framework to judge the quality of each rule. The generated rules are then used in intrusion detection system to detect and to classify network intrusions efficiently in a real-time environment.

But, some limitations of their implemented method are observed. First, the generated rules were partial to the training dataset. Second, though the support-confidence framework is simple to implement and provides improved accuracy to final rules, it requires the whole training datasets to be loaded into memory before any computation. For large training datasets, it is neither efficient nor feasible.

In [12] Hoque, Mukit and Bikas presented an implementation of Intrusion Detection System by applying the theory of genetic algorithm to efficiently detect various types of network intrusive activities. To apply and measure the efficiency of their system they used the standard KDD 99 intrusion detection benchmark dataset and obtained realistic detection rate. To measure the fitness of a chromosome they used the standard deviation equation with distance.

But their performance of detection rate was poor and they failed to reduce the false positive rate in Intrusion Detection System.

## 5. CONCLUSION

In this paper, I have described an overview of some of the current and past intrusion detection technologies which are being utilized for the detection of intrusive activities against computer systems or networks. The different detection challenges that affect the decision policy of the IDS employed in an organization are clearly outlined. I propose to use the new definition of the complement of fuzzy sets where the fuzzy membership value and fuzzy membership function for the complement of a fuzzy set are two different concepts because the surface value is not always counted from the ground level. This new definition of fuzzy sets can classify efficient rule sets. This would help in reducing the false alarm rate occurred in intrusion detection system.

## ACKNOWLEDGEMENTS

I would like to extend my sincere thanks and gratefulness to H.K. Baruah, Professor, Department of Statistics, Gauhati University, Guwahati, India for his kind help, moral support and guidance in preparing this article.

## REFERENCES

[1] J. Gomez & D. Dasgupta, (2002) "Evolving Fuzzy Classifiers for Intrusion Detection", IEEE Proceedings of the IEEE Workshop on Information Assurance, United States Military Academy, West Point, NY.
[2] R. H. Gong, M. Zulkernine & P. Abolmaesumi, (2005) "A Software Implementation of a Genetic Algorithm Based Approach to Network Intrusion Detection", Proceedings of the Sixth International Conference on Software Engineering, Artificial Intelligence, Networking and Parallel/Distributed Computing and First ACIS International Workshop on Self-Assembling Wireless Networks.
[3] T. Xia, G. Qu, S. Hariri & M. Yousif, (2005) "An Efficient Network Intrusion Detection Method Based on Information Theory and Genetic Algorithm", Proceedings of the 24th IEEE International Performance Computing and Communications Conference, Phoenix, AZ, USA.
[4] Yao, J. T., S.L. Zhao & L.V. Saxton, (2005) "A Study On Fuzzy Intrusion Detection", In Proceedings of the Data Mining, Intrusion Detection, Information Assurance, And Data Networks Security, SPIE, Vol. 5812, Orlando, Florida, USA , pp. 23-30.
[5] B. Abdullah, I. Abd-alghafar, Gouda I. Salama & A. Abd-alhafez, (2009) "Performance Evaluation of a Genetic Algorithm Based Approach to Network Intrusion Detection System", 13th International Conference on Aerospace Sciences and Aviation Technology (ASAT), May 26-28.
[6] Hemanta K. Baruah, (2011) "Towards Forming A Field Of Fuzzy Sets", International Journal of Energy, Information and Communications (IJEIC), Vol. 2, Issue 1, February, pp. 16-20.
[7] Hemanta K. Baruah, (2011) "The Theory of Fuzzy Sets: Beliefs and Realities", International Journal of Energy, Information and Communications (IJEIC), Vol. 2, Issue 2, pp. 1-22.
[8] Tridiv Jyoti Neog & Dushmanta Kumar Sut, (2011) "Complement of an Extended Fuzzy Set", International Journal of Computer Applications (IJCA), Vol. 29-No.3, September, pp. 39-45.
[9] Zadeh L A, (1965) "Fuzzy Sets", Information and Control, Vol.8, pp. 338-353.
[10] Y. Dhanalakshmi & Dr. I. Ramesh Babu, (2008) "Intrusion Detection Using Data Mining Along Fuzzy Logic and Genetic Algorithms", International Journal of Computer Science and Network Security (IJCSNS), Vol.8, No.2, February, pp. 27-32.

[11] W. Lu & I. Traore, (2004) "Detecting New Forms of Network Intrusion Using Genetic Programming", Computational Intelligence, vol. 20, pp. 3, Blackwell Publishing, Malden, pp. 475-494.
[12] Mohammad Sazzadul Hoque, Md. Abdul Mukit & Md. Abu Naser Bikas, (2012) "An Implementation of Intrusion Detection System using Genetic Algorithm", International Journal of Network Security and Its Applications (IJNSA),Vol.4, No.2, March, pp. 109-120.
[13] W. Li, (2004) "A Genetic Algorithm Approach to Network Intrusion Detection", SANS Institute, USA.
[14] A. Sung & S. Mukkamala, (2003) "Identifying important features for intrusion detection using support vector machines and neural networks", in Symposium on Applications and the Internet, pp. 209– 216.
[15] J. P. Planquart, "Application of Neural Networks to Intrusion Detection", SANS Institute Reading Room.
[16] R. G. Bace, (2000) "Intrusion Detection", Macmillan Technical Publishing.
[17] S. Kumar & E. Spafford, (1995) "A Software architecture to Support Misuse Intrusion Detection", in the 18th National Information Security Conference, pp. 194-204.
[18] K. Ilgun, R. Kemmerer & P. A. Porras, (1995) "State Transition Analysis: A Rule-Based Intrusion Detection Approach", IEEE Transaction on Software Engineering, pp. 181-199.
[19] S. Kumar, (1995) "Classification and Detection of Computer Intrusions", Purdue University.
[20] V. Bobor, (2006) "Efficient Intrusion Detection System Architecture Based on Neural Networks and Genetic Algorithms", Department of Computer and Systems Sciences, Stockholm University/Royal Institute of Technology, KTH/DSV.
[21] KDD-CUP (99) Task Description; http://kdd.ics.uci.edu/databases/kddcup99/task.html
[22] KDD Cup (1999): Tasks; http://www.kdd.org/kddcup/index.php?section=1999&method=task
[23] KDD Cup (1999): Data; http://www.kdd.org/kddcup/index.php?section=1999&method=data
[24] H. G. Kayacık, A. N. Zincir-Heywood & M. I. Heywood, (2005) "Selecting Features for Intrusion Detection: A Feature Relevance Analysis on KDD 99 Intrusion Detection Datasets".
[25] G. Folino, C. Pizzuti & G. Spezzano, (2005) "GP Ensemble for Distributed Intrusion Detection Systems", ICAPR, pp. 54-62.

**Authors**

**Mostaque Md. Morshedur Hassan**

Mostaque Md. Morshedur Hassan is a senior Assistant Professor of Computer Science in the department of Computer Science and Information Technology at Lalit Chandra Bharali College (LCBC), Maligaon, Guwahati, Assam, India. He holds his Master of Computer Application (MCA) degree from Allahabad Agricultural Institute Deemed University, Allahabad. His area of interests includes Network Security, Intrusion Detection and Prevention, Wireless Security, Web Security, Fuzzy Logic, and Social Networking Site.

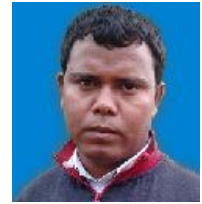